\documentclass[amsmath,amssymb,twocolumn,prc,showpacs,nofootinbib,floatfix,superscriptaddress]{revtex4-1}
\usepackage{graphics}
\usepackage{dcolumn}
\usepackage{bm}
\usepackage{longtable}
\usepackage[dvips]{graphicx}
\usepackage{dcolumn}
\usepackage{footnote}
\usepackage{amssymb}
\usepackage{multirow}
\usepackage[hidelinks]{hyperref}
\usepackage{xcolor}
\hypersetup{
    colorlinks,
    linkcolor={red!50!black},
    citecolor={blue!50!black},
    urlcolor={blue!80!black}
}
\begin{document}

\title{Doppler Broadening in $^{20}$Mg($\beta p\gamma$)$^{19}$Ne Decay}
\author{B. E. Glassman}
\email{glassm16@msu.edu}
\affiliation{Department of Physics and Astronomy, Michigan State University, East Lansing, Michigan 48824, USA}
\affiliation{National Superconducting Cyclotron Laboratory, Michigan State University, East Lansing, Michigan 48824, USA}
\author{D. P\'{e}rez-Loureiro}
\email{david.perez.loureiro@gmail.com}
\affiliation{National Superconducting Cyclotron Laboratory, Michigan State University, East Lansing, Michigan 48824, USA}
\author{C. Wrede}
\email{wrede@nscl.msu.edu}
\affiliation{Department of Physics and Astronomy, Michigan State University, East Lansing, Michigan 48824, USA}
\affiliation{National Superconducting Cyclotron Laboratory, Michigan State University, East Lansing, Michigan 48824, USA}
\author{J. Allen}
\affiliation{Department of Physics, University of Notre Dame, Notre Dame, Indiana 46556, USA}
\author{D. W. Bardayan}
\affiliation{Department of Physics, University of Notre Dame, Notre Dame, Indiana 46556, USA}
\author{M. B. Bennett}
\affiliation{Department of Physics and Astronomy, Michigan State University, East Lansing, Michigan 48824, USA}
\affiliation{National Superconducting Cyclotron Laboratory, Michigan State University, East Lansing, Michigan 48824, USA}
\author{K. A. Chipps}
\affiliation{Oak Ridge National Laboratory, Oak Ridge, Tennessee 37831, USA}
\affiliation{Department of Physics and Astronomy, University of Tennesssee, Knoxville, Tennessee 37996, USA}
\author{M. Febbraro}
\affiliation{Oak Ridge National Laboratory, Oak Ridge, TN 37831, USA}
\affiliation{Department of Physics and Astronomy, University of Tennesssee, Knoxville, Tennessee 37996, USA}
\author{M. Friedman}
\affiliation{National Superconducting Cyclotron Laboratory, Michigan State University, East Lansing, Michigan 48824, USA}
\author{C. Fry}
\affiliation{Department of Physics and Astronomy, Michigan State University, East Lansing, Michigan 48824, USA}
\affiliation{National Superconducting Cyclotron Laboratory, Michigan State University, East Lansing, Michigan 48824, USA}
\author{M.~R. Hall}
\affiliation{Department of Physics, University of Notre Dame, Notre Dame, Indiana 46556, USA}
\author{O. Hall}
\affiliation{Department of Physics, University of Notre Dame, Notre Dame, Indiana 46556, USA}
\author{S. N. Liddick}
\affiliation{Department of Chemistry, Michigan State University, East Lansing, Michigan 48824, USA}
\affiliation{National Superconducting Cyclotron Laboratory, Michigan State University, East Lansing, Michigan 48824, USA}
\author{P.~O'Malley}
\affiliation{Department of Physics, University of Notre Dame, Notre Dame, Indiana 46556, USA}
\author{W. -J. Ong}
\affiliation{Department of Physics and Astronomy, Michigan State University, East Lansing, Michigan 48824, USA}
\affiliation{National Superconducting Cyclotron Laboratory, Michigan State University, East Lansing, Michigan 48824, USA}
\author{S. D. Pain}
\affiliation{Oak Ridge National Laboratory, Oak Ridge, Tennessee 37831, USA}
%
%
\author{S. B. Schwartz}
\affiliation{Department of Physics and Astronomy, Michigan State University, East Lansing, Michigan 48824, USA}
\affiliation{National Superconducting Cyclotron Laboratory, Michigan State University, East Lansing, Michigan 48824, USA}
\author{P. Shidling}
\affiliation{Cyclotron Institute, Texas A \& M University College Station, Texas 77843, USA}
\author{H. Sims}
\affiliation{University of Surrey, GU2 7XH, Guildford, UK}
\author{P. Thompson}
\affiliation{Oak Ridge National Laboratory, Oak Ridge, Tennessee 37831, USA}
\affiliation{Department of Physics and Astronomy, University of Tennesssee, Knoxville, Tennessee 37996, USA}
\author{H. Zhang}
\affiliation{Department of Physics and Astronomy, Michigan State University, East Lansing, Michigan 48824, USA}
\affiliation{National Superconducting Cyclotron Laboratory, Michigan State University, East Lansing, Michigan 48824, USA}
\date{\today}

\begin{abstract}
\noindent
\textbf{Background}: The $^{15}$O($\alpha ,\gamma$)$^{19}$Ne bottleneck reaction in Type I x-ray bursts is the most important thermonuclear reaction rate to constrain experimentally, in order to improve the accuracy of burst light-curve simulations. A proposed technique to determine the thermonuclear rate of this reaction employs the $^{20}$Mg($\beta p\alpha$)$^{15}$O decay sequence. The key $^{15}$O($\alpha ,\gamma$)$^{19}$Ne resonance at an excitation of 4.03 MeV is now known to be fed in $^{20}$Mg($\beta p\gamma$)$^{19}$Ne; however, the energies of the protons feeding the 4.03 MeV state are unknown. Knowledge of the proton energies will facilitate future $^{20}$Mg($\beta p \alpha$)$^{15}$O measurements.

\noindent
\textbf{Purpose}: To determine the energy of the proton transition feeding the 4.03 MeV state in $^{19}$Ne.

\noindent
\textbf{Method}: A fast beam of $^{20}$Mg was implanted into a plastic scintillator, which was used to detect $\beta$ particles. 16 high purity germanium detectors were used to detect $\gamma$ rays emitted following $\beta p$ decay. A Monte Carlo method was used to simulate the Doppler broadening of $^{19}$Ne $\gamma$ rays and compare to the experimental data.

\noindent
\textbf{Results}: The center of mass energy between the proton and $^{19}$Ne, feeding the 4.03 MeV state, is measured to be 1.21${^{+0.25}_{-0.22}}$ MeV, corresponding to a $^{20}$Na excitation energy of 7.44${^{+0.25}_{-0.22}}$ MeV. Absolute feeding intensities and $\gamma$-decay branching ratios of $^{19}$Ne states were determined including the 1615 keV state, which has not been observed before in this decay. A new $\gamma$ decay branch from the 1536 keV state in $^{19}$Ne to the ground state is reported. The lifetime of the 1507 keV state in $^{19}$Ne is measured to be 4.3${^{+1.3}_{-1.1}}$ ps resolving discrepancies in the literature. Conflicting $^{20}$Mg($\beta p$) decay schemes in published literature are clarified. 

\noindent
\textbf{Conclusions}: The utility of this Doppler broadening technique to provide information on $\beta$-delayed nucleon emission and excited-state lifetimes has been further demonstrated. In particular, knowledge of the proton energies feeding the 4.03 MeV $^{19}$Ne state in $^{20}$Mg $\beta$ decay will facilitate future measurements of the $\alpha$-particle branching ratio.

\end{abstract}


\maketitle

\section{Introduction}
A Type I x-ray burst can occur when a binary star system, consisting of a neutron star and hydrogen-rich star, is close enough that matter from the hydrogen-rich star is accreted onto the surface of the neutron star \cite{Lewin1993}. The increasing heat from dense accumulated matter on the surface of the neutron star can lead to thermonuclear runaway and is a likely site of the rapid proton capture process \cite{Wallace1981}, synthesizing new elements up to mass number $A\simeq100$ \cite{Schatz2006}.

In the study of Type I x-ray bursts there are a few reaction bottlenecks whose unknown or highly uncertain rates can have large effects on simulated burst profiles. The most important reaction rate to determine is the $^{15}$O($\alpha,\gamma$)$^{19}$Ne Hot CNO cycle breakout reaction which heavily affects the onset of the burst \cite{Cyburt2016}. A single resonance is expected to dominate the reaction rate and corresponds to an excitation energy of $E_x$($^{19}$Ne) = 4.03 MeV. 

It is not possible with current facilities to directly measure the $^{15}$O($\alpha,\gamma$)$^{19}$Ne reaction rate because an $^{15}$O rare isotope beam of sufficient intensity is not available. However, the resonance strength can be indirectly determined from measurements of the spin, lifetime, and branching ratio $\Gamma_\alpha/\Gamma$ of the 4.03 MeV state. Currently, the spin is known to be $3/2^{+}$ \cite{Tilley1995} and the lifetime has been measured \cite{Kanungo2006,Tan2005,Mythili2008} to sufficient precision, while only sensitive upper limits have been placed on the $\alpha$-decay branch\cite{Davids2011,Rehm2003,Davids2003,Tan2007}.

A recently proposed technique for measuring  $\Gamma_\alpha/\Gamma$ employs the decay sequence $^{20}$Mg($\beta p\alpha$) for which the last step is $\alpha$-particle emission to $^{15}$O, the inverse of $\alpha$ capture \cite{Wrede2017}. An important component to identifying $p-\alpha$ coincidence events of interest is the unknown energy of the proton(s) emitted from the excited states in $^{20}$Na which feed the 4.03 MeV state in $^{19}$Ne.

In the present work, we employ a Doppler-broadening technique to measure the proton energy. When a nucleon is emitted from a nucleus, following $\beta$ decay, the momentum of the system must be conserved so the daughter nucleus will recoil with equal and opposite momentum as the ejected nucleon. If a $\gamma$ ray is emitted before an excited daughter nucleus has time to stop the resulting $\gamma$ ray will be Doppler shifted. The resulting $\gamma$ ray line shape will be broadened. The broadened feature preserves information about the energies of the emitted nucleons, which is modeled using a Monte Carlo simulation method. 

This Doppler broadening method was first used in order to study $^{11}$Li($\beta n\gamma$)$^{10}$Be where direct measurement of neutrons is very difficult but the relatively light nuclei provide substantial recoil velocities \cite{Fynbo2004,Sarazin2004,Mattoon2009}. The analysis of the line-shape allowed for the construction of a partial decay scheme as well as the ability to measure nuclear lifetimes of the $\gamma$-decaying states. This method was recently utilized in $^{26}$P($\beta p\gamma$)$^{25}$Al to extend the method to higher masses and apply it to proton emission for the first time \cite{Schwartz2015}.

In previous measurements, excited state energies, lifetimes, and $\gamma$-branching ratios of $^{19}$Ne levels have been measured for nearly all states known to be fed by $^{20}$Mg($\beta p\gamma$)  \cite{Tilley1995,Tan2005,Kanungo2006,Gill1970}. Previous experiments have measured protons from the $^{20}$Mg($\beta p$)$^{19}$Ne decay sequence directly \cite{Piechaczek1995,Lund2016,Sun2017}; however, some aspects of the various decay schemes constructed are inconsistent. Additionally there are excited states in $^{19}$Ne with no proton feeding information available. In the present work, we use the Doppler broadening line shape analysis method to resolve some of these inconsistencies and provide new information about states with unknown proton feedings or lifetimes.

\section{Experimental Setup}

\begin{figure}
\includegraphics[width=0.5\textwidth]{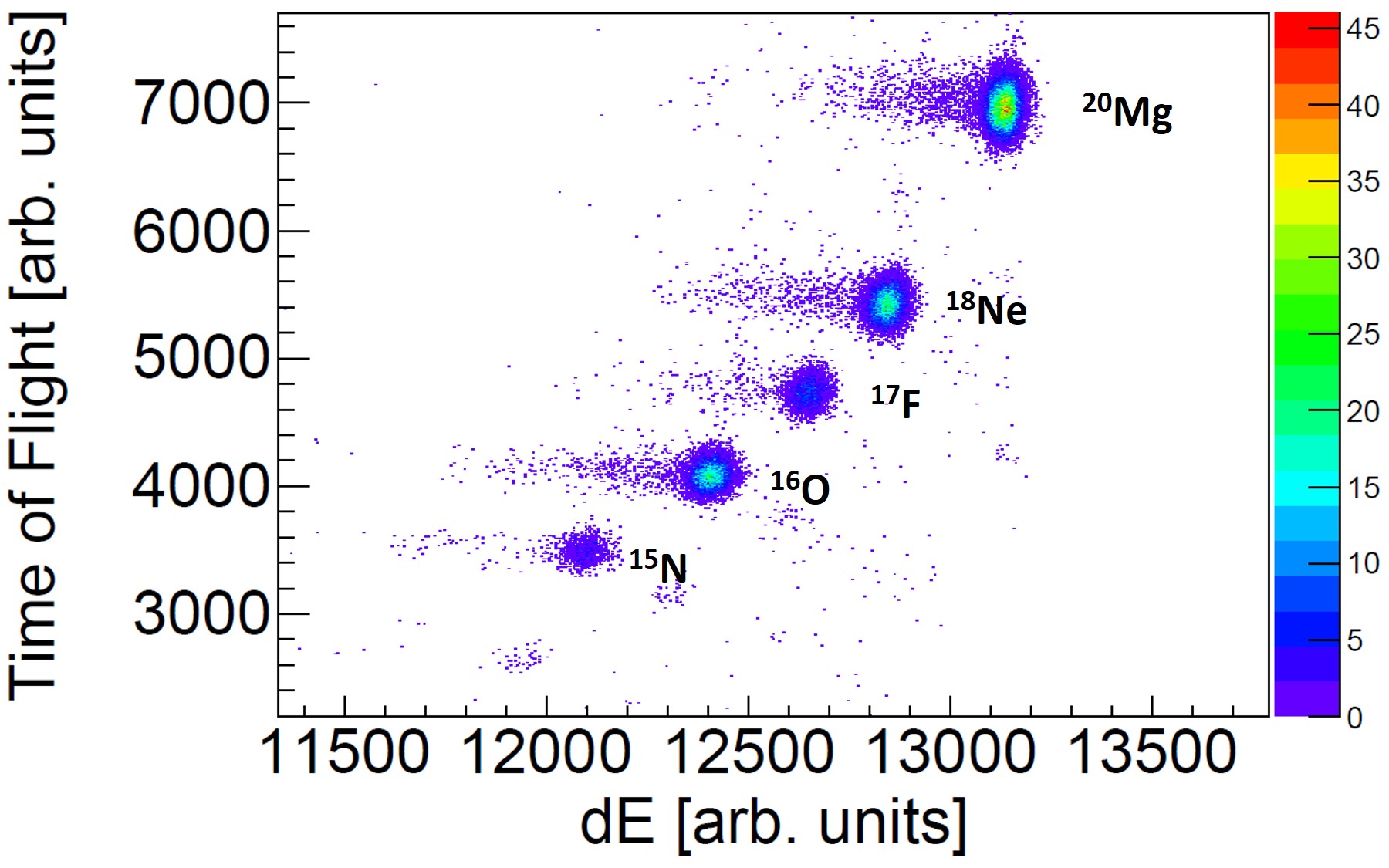}
\caption{Particle identification plot obtained with an attenuated beam in between production runs. The time of flight was determined over a 25 m path between the scintillator placed at the focal plane of the A1900 and the Si PIN detector. The energy loss dE was gathered from the energy deposited in the PIN detector.}
\label{pid}
\end{figure}

The  $^{20}$Mg $\beta$-decay experiment was performed at the National Superconducting Cyclotron Laboratory (NSCL) and has been described in Refs. 
\cite{Glassman2018-2,Wrede2017,Glassman2015}. Briefly, a $^{24}$Mg primary beam was accelerated by the K500 and K1200 coupled cyclotrons to 170 MeV/u, and impinged on a 961 mg/cm$^2$ $^9$Be transmission target. The fast secondary beam contained the desired $^{20}$Mg as well as other fragments. Fragments whose momentum to charge ratio differed from $^{20}$Mg were removed from the beam using the A1900 magnetic fragment separator, and similarly a bulk of the fragments with $Z\neq12$  were removed using a 594 mg/cm$^2$ Al wedge \cite{Morrissey2003}. A 300-$\mu$m-thick Si transmission detector was lowered periodically into the beamline to determine the beam composition, $\approx$1 m upstream of the experimental setup, using the $\Delta$E-TOF method. The time-of-flight was measured over a path of 25 m using  the Si detector and a scintillator at the focal plane of the A1900 (Fig. \ref{pid}). The final beam consisted of 34\% $^{20}$Mg ($Q_{EC}$=10.7 MeV), 24\% $^{18}$Ne  ($Q_{EC}$=4.4 MeV), 12\% $^{17}$F  ($Q_{EC}$=2.8 MeV), 22\% $^{16}$O (stable) and 8\% $^{15}$N (stable). 

 Up to 4000 $^{20}$Mg ions s$^{-1}$ were delivered to the experimental setup and were implanted in a 5 cm $\times$ 5 cm $\times$ 2.5cm thick plastic scintillator, which detected ion implantations and $\beta$-decays. The scintillator was surrounded by two rings of eight segmented HPGe (high-purity germanium detectors) each, the Segmented Germanium Array (SeGA), which detected $\gamma$ rays. Data from these $\beta$-decay events were collected by the NSCL Digital Data Acquisition System \cite{Prokop2014}.

\section{Data and Analysis}

We produce a total $\beta$-particle gated $\gamma$-ray spectrum to compare to Doppler broadening simulations. Three of the SeGA detectors had resolutions that were 25-110\% worse than average and would increase systematic errors in the Doppler broadening analysis, and since we did not suffer significantly by losing these statistics, only 13 of the 16 detectors were used to analyze most peaks. The resulting spectra from these 13 SeGA detectors were added to produce a total $\beta$-gated $\gamma$ ray spectrum \cite{Wrede2017,Glassman2018-2}.

\subsection{$\gamma$-ray energy and efficiency calibration}

To create the total spectrum, the $\gamma$-ray energy spectrum of each SeGA detector was linearly gain-matched run by run using room background lines at 2614.511 $\pm$ 0.010 keV and 1460.820 $\pm$ 0.005 keV from the $\beta$-decays $^{208}$Tl and $^{40}$K respectively. An exponentially modified gaussian function, having the form (Eq. \ref{eq:emg}), was used to model the response function for each SeGA detector and the maximum value from the fit of these $\gamma$ ray lines was used for linear gain-matching. The exponentially modified gaussian is characterized by an exponential parameter $\lambda$, width $\sigma$, peak position $\mu$, energy $x$, and normalization $N$.

$\gamma$ ray energies were calibrated using well known room background  $\gamma$ ray lines from decays of $^{40}$K and $^{208}$Tl in addition to strong $\gamma$ ray lines from $^{20}$Na($\beta\gamma$) and $^{20}$Na($\beta\alpha\gamma$)  at energies 1633.602 $\pm$ 0.015, 3332.54 $\pm$ 0.20, 6128.63 $\pm$ 0.04, 8237 $\pm$ 4, and 8638 $\pm$ 3 keV. $\gamma$-ray energies are reported in the lab frame and excitation energies are reported with recoil corrections applied.


\begin{multline}
f(x;N,\mu,\sigma,\lambda) = \dfrac{N\sigma}{\lambda}\sqrt{\dfrac{\pi}{2}}exp\Bigg(\dfrac{1}{2}\Big(\dfrac{\sigma}{\lambda}\Big)^2 +\dfrac{x-\mu}{\lambda}\Bigg)\\erfc\Bigg(\dfrac{1}{\sqrt{2}}\Big(\dfrac{\sigma}{\lambda}+\dfrac{x-\mu}{\sigma}\Big)\Bigg)
\label{eq:emg}
\end{multline}

\begin{figure}
\includegraphics[width=0.5\textwidth]{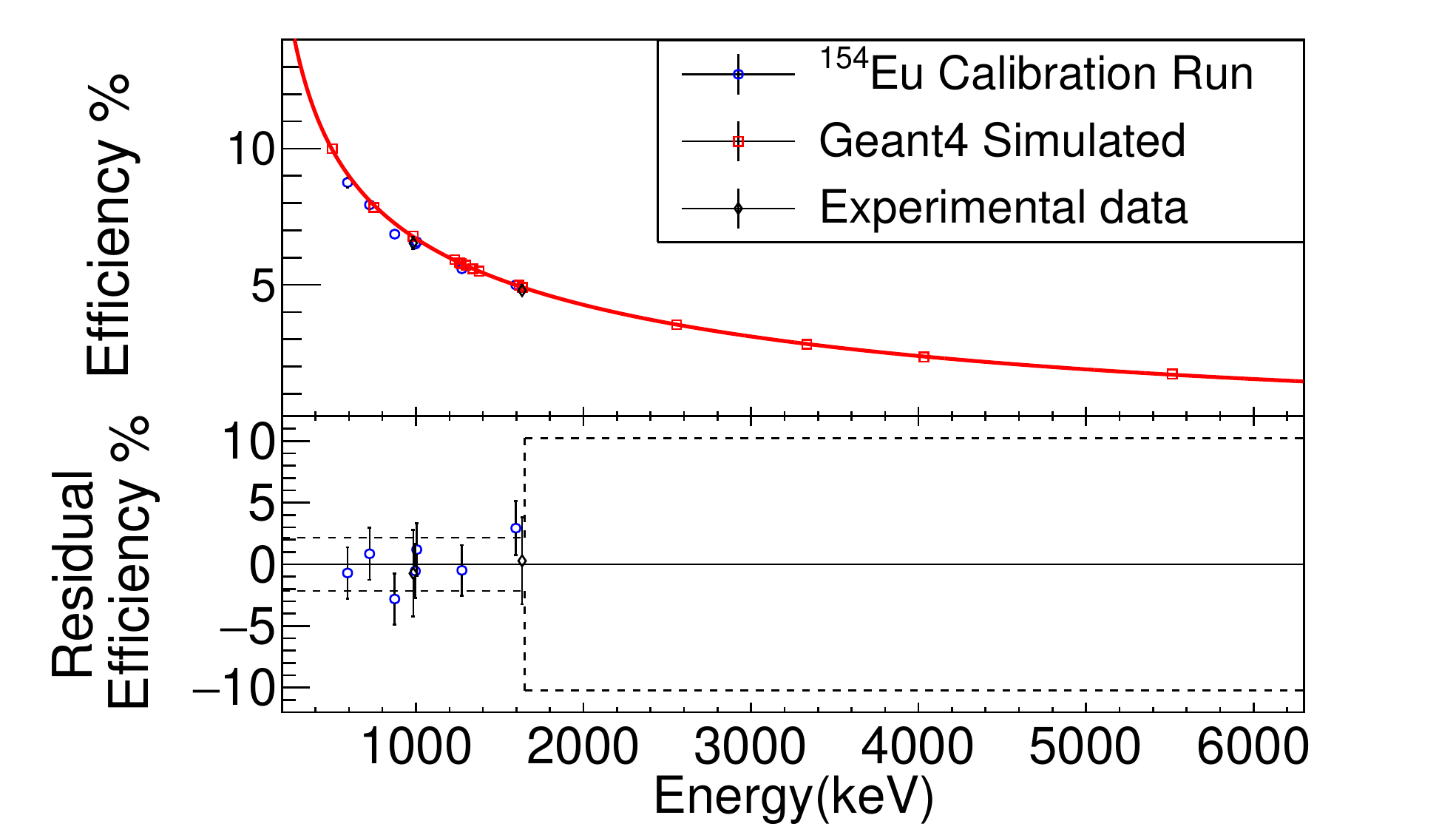}
\caption{Upper Panel: Absolute $\gamma$ ray photopeak efficiency curve (solid [red online] line) for all 16 SeGA detectors generated by scaling { \sc Geant}4 simulated efficiencies over a range of energies by a constant factor of 0.975 to match the radioactive source calibration efficiency. Lower Panel: Residuals between the calibration source efficiency and the scaled fit of the { \sc Geant}4 simulated efficiencies. The dotted lines represent a one standard deviation uncertainty envelope.}
\label{effplotg}
\end{figure}

An absolutely calibrated source of $^{154}$Eu was used to determine the $\gamma$ ray efficiency (Fig. \ref{effplotg}). This provided us with efficiency data points ranging from 123.1 keV to 1596.4 keV. The geometry of our experimental setup was used as input for a { \sc Geant}4 Monte Carlo simulation to determine a $\gamma$ ray photopeak efficiency curve for all 16 SeGA detectors, which matched with our calibration source efficiency when multiplied by a constant factor of 0.975. Using this simulated efficiency curve we extrapolated the efficiency to higher energies. This procedure was shown to be accurate in previous experiments with very similar geometries and calibration peaks that spanned a wider range of energies \cite{Bennett2018,PrezLoureiro2016}. Below 1600 keV, we interpolated the efficiency, so a flat statistical uncertainty of 0.8\% is used. An additional 2\% systematic uncertainty is applied to account for summing of $\gamma$ rays from the calibration source.   Previous experiments used a 5\% uncertainty for efficiencies determined using {\sc Geant4} to extrapolate to higher energies \cite{Bennett2018,PrezLoureiro2016}, however, these experiments had measured efficiencies up to 5 MeV and 2.8 MeV. Since we must extrapolate as high as 4.03 MeV from measured efficiencies only up to 1.6 MeV, a more conservative systematic uncertainty of 10\% is adopted for the extrapolated $\gamma$ ray efficiency at 4.03 MeV. This value is somewhat arbitrary but certainly conservative.

\subsection{$\beta$ decay detection efficiency}
To reduce background in the $\beta$-particle gated $\gamma$ ray spectrum, a gate on the scintillator energy was applied, which differentiated $\beta$-decay events and the much higher-energy ion implantation events. A 1 $\mu$s timegate was also applied to reduce the contribution from random coincidences. The well known 984 keV $\gamma$ ray from $^{20}$Mg($\beta\gamma$)$^{20}$Na decay, with branching ratio 0.697 $\pm$ 0.012 \cite{Piechaczek1995}, is used to normalize the number of $^{20}$Mg $\beta$ decays occuring in our experiment. A branching ratio of 0.725 $\pm$ 0.025 was also measured in a recent experiment \cite{Lund2016}, in agreement with the more precise value. Therefore, it is important to characterize our scintillator's $\beta$-decay detection efficiency as a function of $\beta$-endpoint energy and proton energy.
 
 \begin{figure}
\includegraphics[width=0.5\textwidth]{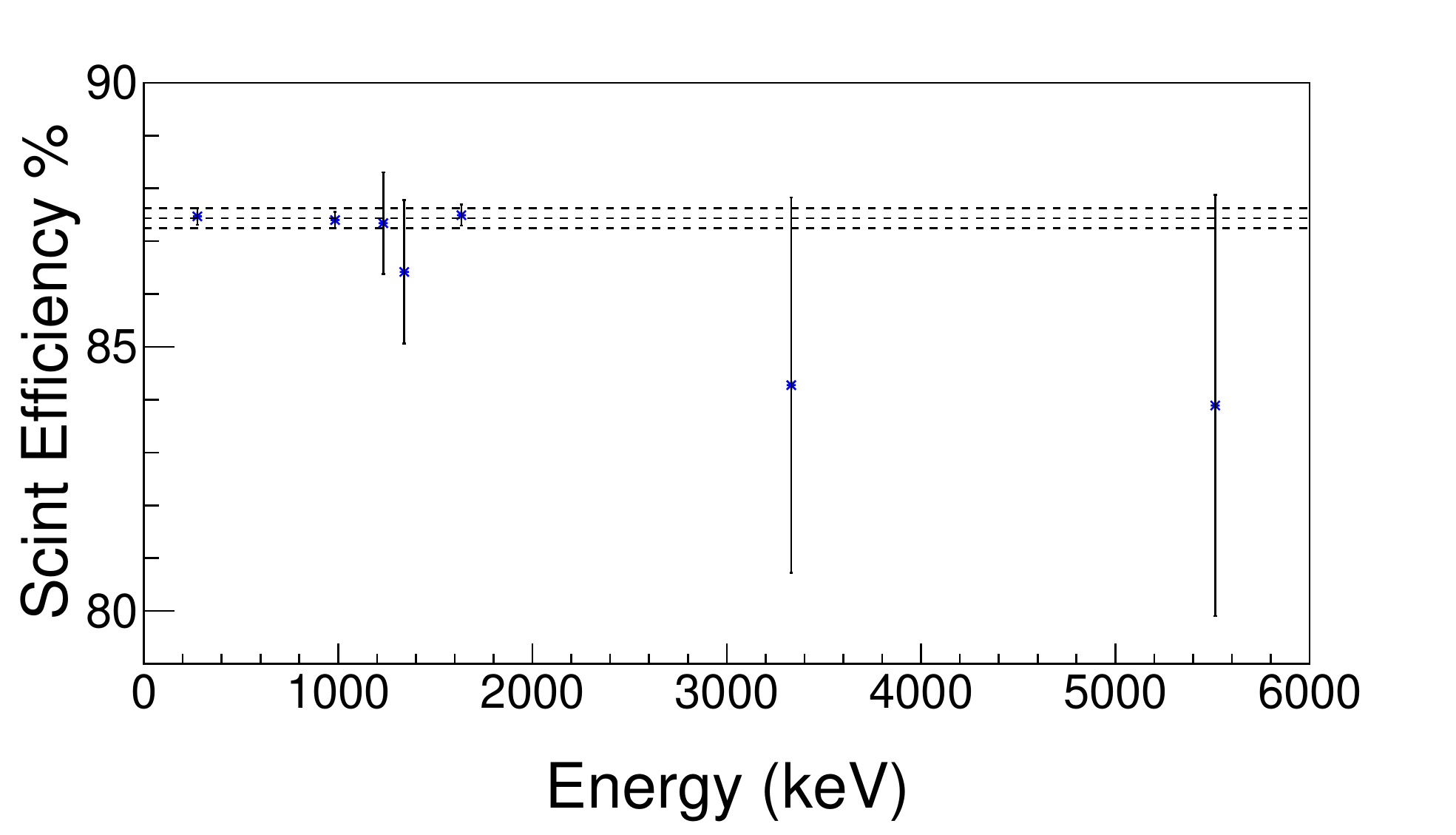}
\caption{Ratio of counts in photopeaks of scintillator-gated $\gamma$ ray spectrum to counts in ungated spectrum. The mean of the measurements at 87.4\% is denoted by the central dashed line with a 1 standard deviation envelope of 0.2\%. Ratios are measured for $^{20}$Mg($\beta\gamma$)$^{20}$Na, $^{20}$Na($\beta\gamma$)$^{20}$Ne and $^{20}$Mg($\beta p\gamma$)$^{19}$Ne peaks. }
\label{effplots}
\end{figure}
The scintillator efficiency is measured by comparing the number of counts in the ungated $\gamma$ ray spectrum to the number of counts in the scintillator gated $\gamma$ ray spectrum for $^{20}$Mg($\beta\gamma$), $^{20}$Na($\beta\gamma$) and $^{20}$Mg($\beta p\gamma$) peaks. Each peak represents a different sample of $\beta$-endpoint energies and proton energies. A constant $87.4 \pm 0.2$\% scintillator efficiency was consistent over a large sample of $\gamma$ ray peaks (Fig. \ref{effplots}) and therefore the scintillator efficiency to detect $\beta$ decays was assumed to be constant. 

\subsection{Doppler Broadening Analysis}

A Monte Carlo simulation was developed to model the Doppler broadening. Inputs of the simulation include the Center of Mass (CoM) energy between the emitted proton and $^{19}$Ne$^*$ state (denoting the $^{19}$Ne in an excited state), the lifetime and excitation energy of the $^{19}$Ne$^*$ state, the stopping power of the implantation material (Polyvinyltoluene), and the response function of each SeGA detector.

 The recoiling $^{19}$Ne$^*$ is given an initial kinetic energy based on the CoM energy of the proton emission from an excited $^{20}$Na state which is, to a good approximation, at rest in the plastic scintillator. The Monte Carlo simulation works by first assuming a lifetime for a $^{19}$Ne$^*$ state and randomly sampling the exponential decay curve distribution. 

We require an understanding of the stopping power to determine how much kinetic energy the recoiling  $^{19}$Ne$^*$ atom will lose before it emits a $\gamma$ ray. The stopping power is determined as a function of recoil energy using SRIM \cite{Ziegler2010}. Depending on the sampled lifetime, the $^{19}$Ne$^*$ atom will slow down a certain amount or stop completely before emitting a $\gamma$ ray. Therefore, the energy lost in the plastic scintillator by the recoiling $^{19}$Ne$^*$ is calculated recursively to model a continuous energy loss. 

 Angular correlations between protons and $\gamma$ rays can have an effect on the overall line-shape \cite{FYNBO2003}. The direction of the proton, produces a $\gamma$-ray angular distribution described by a linear combination of even Legendre polynomials \cite{Ferguson1974}  in the center of mass frame (Eq. \ref{eq:ang}). 
 \begin{equation}
W(\theta_{cm}) = \sum_{2\kappa }A_\kappa P_\kappa(cos(\theta_{cm}))
\label{eq:ang}
\end{equation}
 The highest order Legendre polynomial for each $\gamma$ ray transition is determined by the spin of the proton-emitting $^{20}$Na state, multipolarity of the $\gamma$ ray transition, angular momentum of emitted proton, and spin of $^{19}$Ne$^*$  \cite{Goldfarb1968,Duray1972} such that 
 $$2\kappa_{\text{max}} \leq \text{min}[(2j_{^{20}\text{Na}}),(2L)_\text{max},(2l)_\text{max},(2j_{^{19}\text{Ne*}}-1)_\text{max}]$$
The spin of $^{20}$Na states is constrained to be $0^+$ and $1^+$ in allowed $^{20}$Mg $\beta$ decay, restricting the angular correlation function to the  $P_0(cos(\theta_{cm}))$ and $P_2(cos(\theta_{cm}))$ terms. A first order assumption is made that the isotropic term ($P_0$) dominates and a $P_2$ term will be added if a good fit can not be achieved with this assumption. 

An angle, dependent on the angular distribution function \ref{eq:ang}, is randomly chosen between the recoiling $^{19}$Ne atom and emitted $\gamma$ ray,  to calculate the Doppler shift at the observation point. This $\gamma$ ray enters a random detector and the known response function of that detector is treated as a probability density function which outputs a final observed energy. An ensemble of such events can be used to construct a simulated peak shape for comparison to the actual data.
 
\begin{figure}
\includegraphics[width=0.5\textwidth]{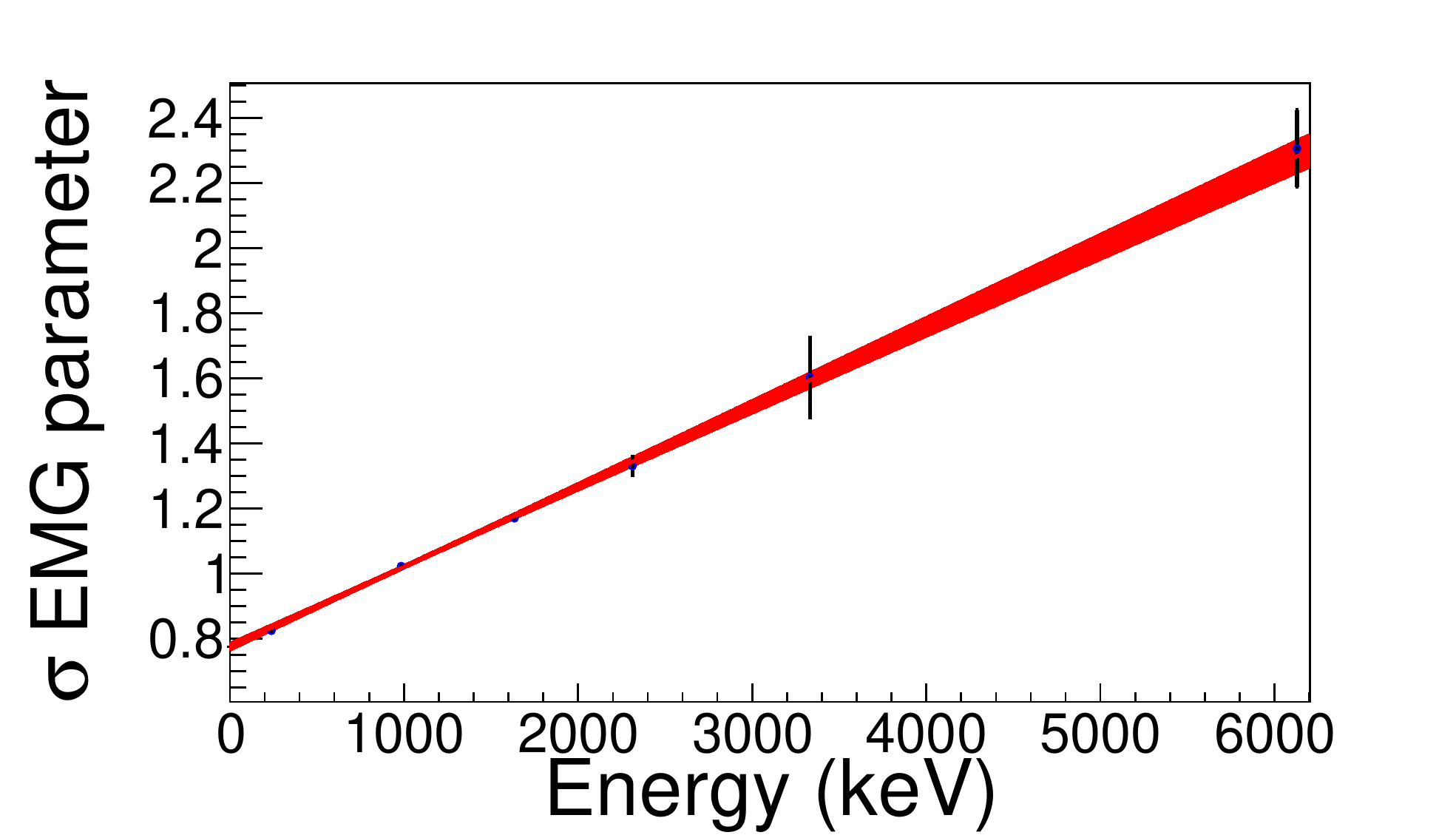}
\caption{An example of the $\sigma$ parameter energy dependence for input to the exponentially-modified-gaussian response function for a single SeGA detector. Each data point corresponds to the value of the $\sigma$ parameter for a particular calibration peak. The $\sigma$ parameter is fit using a line and the confidence band [red online] shows 1 standard deviation uncertainty.}
\label{sigma}
\end{figure}

A response function for each of the SeGA detectors was determined by fitting unbroadened $\beta$-delayed $\gamma$ ray peaks with an exponentially modified Gaussian function at energies 238 ($^{19}$Ne), 984 ($^{20}$Mg), 1634 ($^{20}$Ne), 2312 ($^{14}$N), 3332 ($^{20}$Ne), and 6129 ($^{16}$O) keV (Eq. \ref{eq:emg}). The exponential parameter ($\lambda$) was fixed to 0.7 in order to parameterize $\sigma$ as a function of energy and all other parameters were left free. The value of $\sigma$ was plotted as a function of energy and fit using a linear function (Fig. \ref{sigma}). This parameterization was implemented in the Doppler broadening simulations to mimic the SeGA detectors' response.  Each detector has a slightly different contribution to the total number of counts in the peak depending on efficiency and the simulation reflects this.

\subsection{Doppler Broadening Systematic Uncertainties}
In order to extract accurate information from the Doppler broadening of each peak it is important to first quantify how well we know the inputs and how sensitive the simulation will be to slight changes in each quantity. 

The stopping power, which is determined by SRIM, is expected to be accurate to within ~10\% \cite{Ziegler2010}. The uncertainty in the stopping power is directly related to the uncertainty in the lifetime and will have a greater systematic effect when the lifetime of the excited state is not well known or unknown. 

The exponentially modified gaussian response function is well known for all $^{19}$Ne $\gamma$ ray energies. The $\sigma$ parameter in the response function has \textless0.7\% uncertainty for each detector below 1600 keV, however, this uncertainty is larger in the case of the 4.03 MeV $\gamma$ ray which lies far away from many of the $\beta$-delayed $\gamma$ rays used to model $\sigma$.

The final two inputs of the Doppler broadening simulation, $^{19}$Ne excited state lifetimes and the feeding intensities and energies from $^{20}$Na excited states, can have large literature uncertainties, and in some cases, are unknown. The absolute $^{20}$Mg($\beta p$)$^{19}$Ne$^*$ feeding intensities are obtained from the direct proton measurements of Piechaczek \textit{et al.} and Lund \textit{et al.} \cite{Piechaczek1995,Lund2016} and used when available. The uncertainty in the better known quantity between the lifetime and proton feeding energy is used to determine a systematic uncertainty in measurements of the lesser known quantity, or will be considered a free parameter for $\chi^2$ minimization if there are no prior measurements.

The systematic uncertainties determined from these quantities are combined in quadrature with statistical uncertainties. 

In all fits described below, we are able to achieve a minimum in the $\chi^2_\nu$ distribution close to 1, using an isotropic distribution of $\gamma$-rays with respect to proton distribution. 

\subsection{Background Modeling}
\begin{figure}
\includegraphics[width=0.5\textwidth]{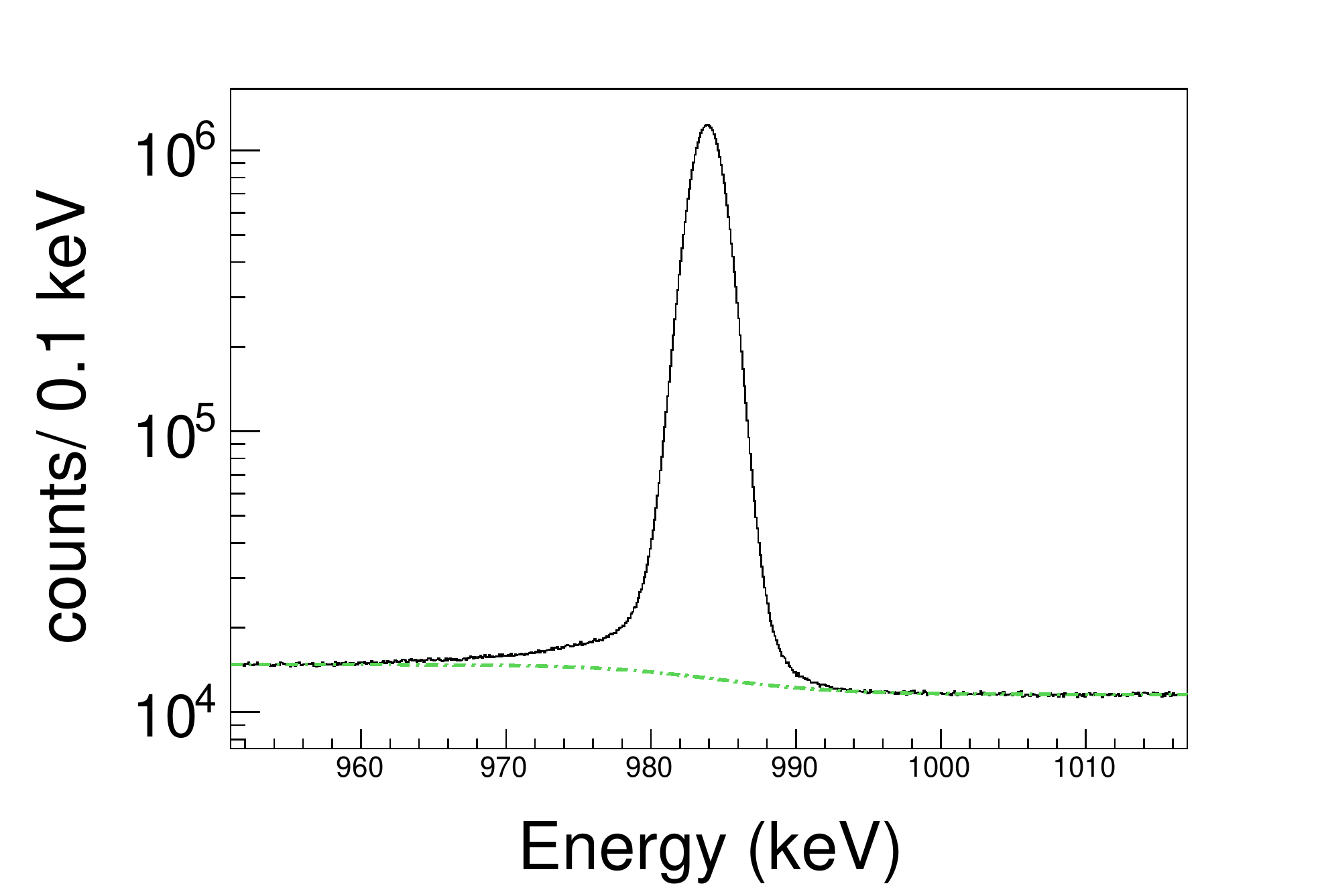}
\caption{(color online) An example background component of a fit function (dot-dashed [green online] line) applied to a high statistics $\gamma$ ray peak (984.25 keV) in the total $\beta$-gated $\gamma$ ray spectrum (solid [black online] line). }
\label{fig:BG}
\end{figure}

To model the background of high-statistics peaks we take a linear fit A on the left side of the peak and linear fit B on the right side of the peak to represent the unique background level on each side and connect them using a continuous step. This was done by weighting function A more heavily to the left of the peak, weighting function B more heavily to the right of the peak, weighting them equally at the maximum of the peak, and summing these functions together to make a tanh-like function. An example of the background model can be seen in Fig. \ref{fig:BG} where the background of the 984 keV peak on the left is significantly higher than on the right due to incomplete charge collection. For very low-statistics peaks a simple linear function is sufficient to model the local background because the step is negligible.

\section{Results and Discussion}

\begin{figure}
\includegraphics[width=0.5\textwidth]{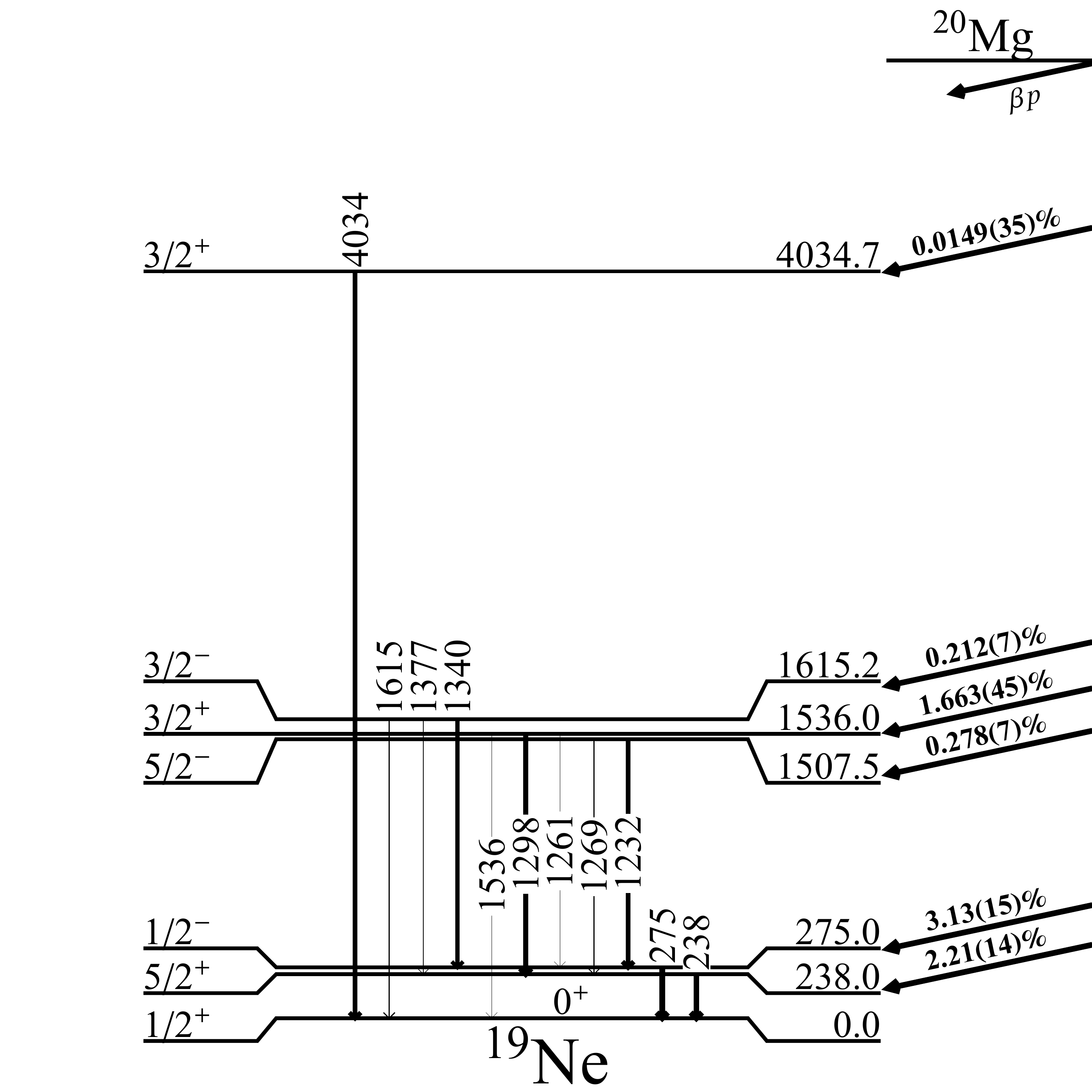}
\caption{$^{19}$Ne level scheme from $^{20}$Mg($\beta p\gamma$)$^{19}$Ne decay deduced from the present work. The $\gamma$ ray transition intensities are denoted by the thicknesses of the arrows, which are proportional to their intensities. The $^{20}$Mg($\beta$p) feeding intensities are denoted by the arrows on the right.}
\label{levscheme}
\end{figure}

The decay scheme presented in Figure \ref{levscheme} is deduced from the $\gamma$ ray spectrum obtained in this experiment. Only the $^{19}$Ne levels which are populated by $^{20}$Mg($\beta$p) are displayed. The measured $^{20}$Mg($\beta$p) intensities and $\gamma$ ray energies are reported in Table \ref{tab:intensities}. The $\gamma$ ray intensities per $^{20}$Mg $\beta$ decay ($I_{\beta p\gamma}$) are determined from the integral of each fit. These values are corrected for the SeGA efficiency and normalized to the number of $^{20}$Mg $\beta$ decays.

We proceed to discuss the individual $^{19}$Ne states.

\begin{table*}
\caption{\label{tab:intensities} Column one reports the $^{19}$Ne excited-state energies populated by $^{20}$Mg($\beta$p), and were determined by applying recoil corrections to the measured $\gamma$ ray energies in the lab-frame (column-four).  Column two reports the measured lifetimes of $^{19}$Ne excited states. Column three reports the intensity of $^{20}$Mg($\beta$p) feedings to each excited state, where each feeding is determined by adding all $\gamma$ ray decays originating from each state and subtracting feeding from higher lying states. Column four reports the measured lab frame energies of each $\gamma$ ray branch. Column five reports the total intensity of each $\gamma$-ray transition per $^{20}$Mg decay. Column six reports the $\gamma$ ray branching ratios for each $^{19}$Ne excited-state. Column seven reports the measured CoM proton energies feeding $^{19}$Ne excited states. }
\begin{ruledtabular}
\begin{tabular}{ c c c c c c c }
 $E_x(^{19}\text{Ne})$ (keV)& $\tau$ (fs) & $I_{^{20}\text{Mg}(\beta p)}$ & $E_\gamma$ (keV) & $I_{\beta p\gamma}$ &  Branch (\%) & $E_{CoM}$ (MeV)\\ 
\hline
  238.04(10) && 0.0221(14) & 238.04(10) & ($3.80 \pm 0.07_{stat} \pm 0.08_{sys}$)$\times10^{-2}$  & 100\\
 274.96(10)  &&   0.0313(15) & 274.96(10) &($3.59 \pm 0.06_{stat} \pm 0.08_{sys}$)$\times10^{-2}$ & 100\\
 1507.52(25)&4.3$^{+1.3}_{-1.1}$&   0.00278(7)  & 1232.49(22) &($2.36 \pm 0.04_{stat} \pm 0.05_{sys}$)$\times10^{-3}$  & 84.9(4)\\
&& & 1269.47(24) &($4.18 \pm 0.12_{stat} \pm 0.09_{sys}$)$\times10^{-4}$ & 15.1(4)\\
  1535.95(24)&&  0.01663(45)& 1260.87(24) & ($6.75 \pm 0.15_{stat} \pm 0.15_{sys}$)$\times10^{-4}$ & 4.05(16)  \\
&&  &1297.94(22) & ($1.539 \pm 0.027_{stat} \pm 0.033_{sys}$)$\times10^{-2}$  & 92.53(35)\\ 
&&  &1535.90(24) & ($5.68 \pm 0.44_{stat} \pm 0.17_{sys}$)$\times10^{-4}$ & 3.42(29)\\ 
  1615.24(30)&&  0.00212(7) &1340.27(25)& ($1.57 \pm 0.03_{stat} \pm 0.03_{sys}$)$\times10^{-3}$ & 74.0(17) & 2.70(23)\\
&&  &1377.1(3)\footnote{Value derived from 238, 275, and 1340 keV $\gamma$ ray peak energies} & ($1.82 \pm 0.41_{stat} \pm 0.04_{sys}$)$\times10^{-4}$& 8.6(18) \\ 
&&  &1615.16(30)\footnote{Value derived from addition of 275 and 1340 keV $\gamma$ ray peak energies} &($3.68 \pm 0.18_{stat} \pm 0.08_{sys}$)$\times10^{-4}$ & 17.4(9)\\ 
  4034.7(16)&&  0.000149(35) & 4034.2(16) & ($1.19 \pm 0.12_{stat} \pm 0.12_{sys}$)$\times10^{-4}$ & 80(15)\footnote{Value adopted from \cite{Tilley1995}} & 1.21${^{+0.25}_{-0.22}}$\\
\end{tabular}
\end{ruledtabular}
\end{table*}

\subsection{$^{19}$Ne 1507 keV $5/2^-$ state}
There are two $\gamma$ rays which are emitted from this state at 1232.5 keV and 1269.3 keV and they are expected to have branching ratios of 88(3)\% and 12(3)\% respectively  \cite{Gill1970}. The 1507 keV excited state lifetime has been previously measured to be $1.4{^{+0.5}_{-0.6}}$ ps \cite{Itahashi1971}, 1.7(3) ps \cite{Tan2005}, and $4.1{^{+3.5}_{-1.4}}$ ps \cite{Gill1970}. Since there is significant tension between the various lifetime measurements, the lifetime was treated as a free paramater for $\chi^2$ minimization.

\begin{figure}
\includegraphics[width=0.5\textwidth]{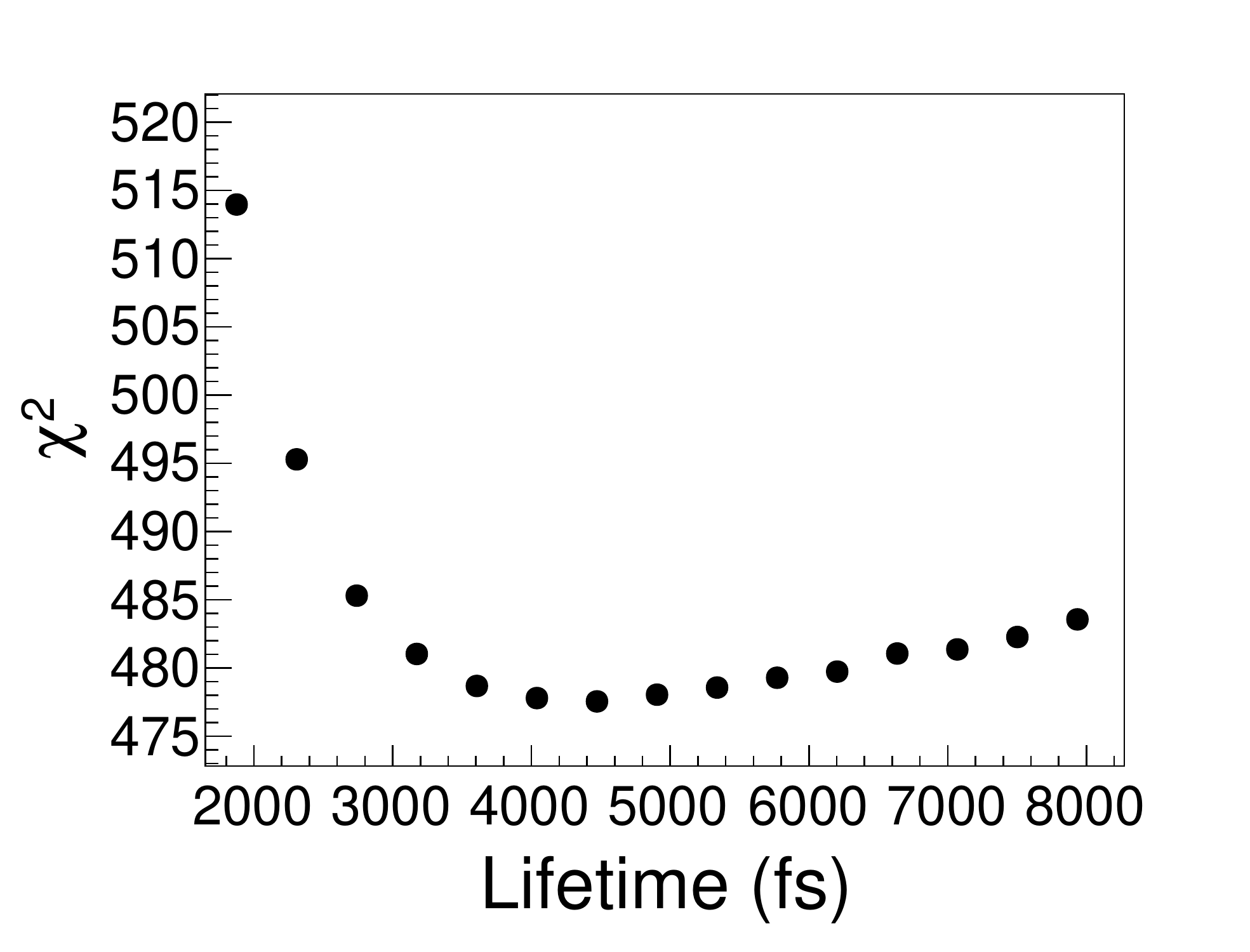}
\caption{$\chi^2$ values determined by simulating  the lifetime of the 1507 keV $^{19}$Ne state for many values and comparing the simulation to the data over 447 degrees of freedom. The minimum is found at 4.3 ps.}
\label{chisq1232}
\end{figure}
\begin{figure}
\includegraphics[width=0.5\textwidth]{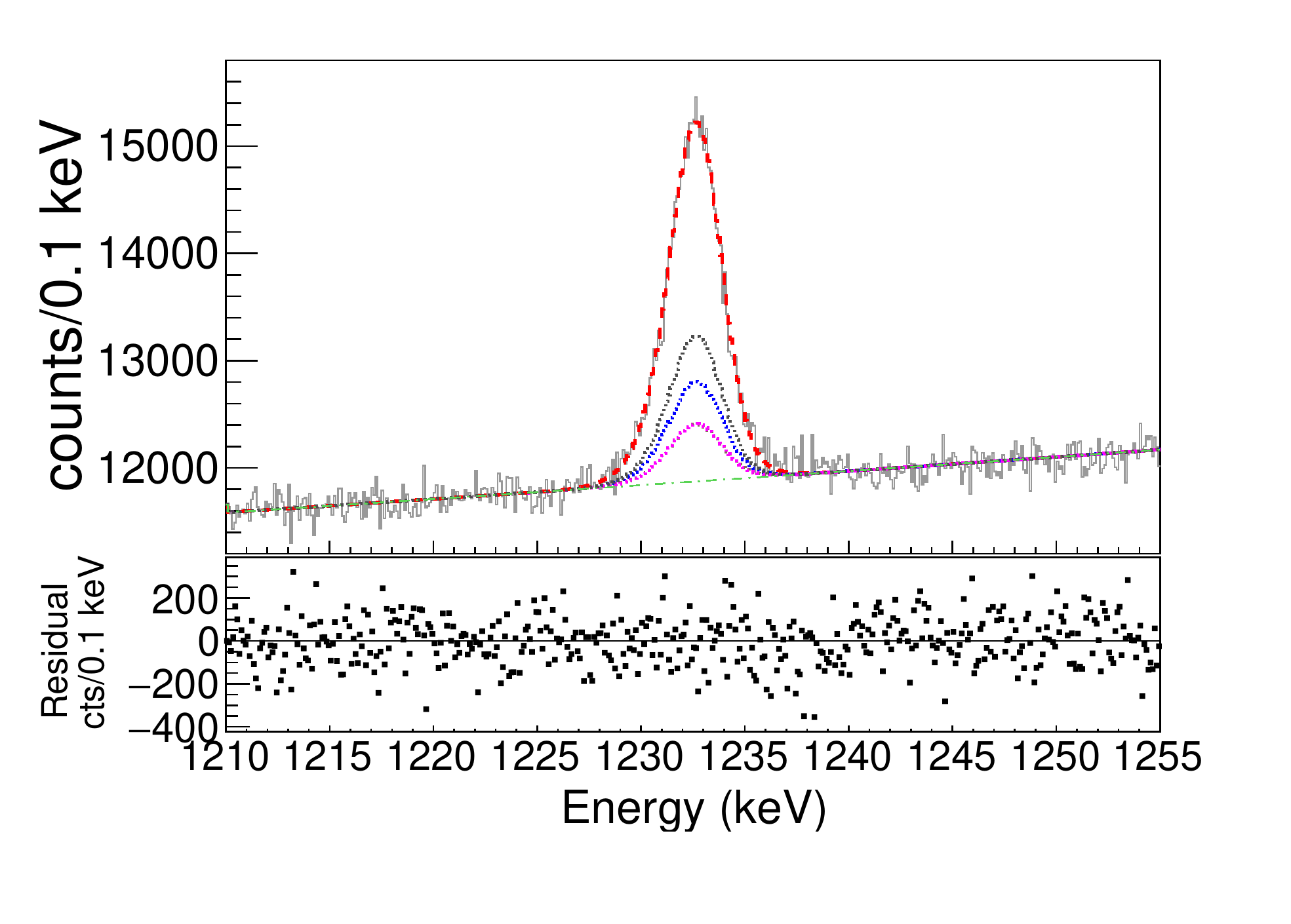}
\caption{(color online) Upper panel: The fit of the 1232 keV $\gamma$ ray peak is produced by using 4.3 ps lifetime as well as proton feeding intensities from Piechaczek \textit{et al.} \cite{Piechaczek1995}. The solid [gray online] line represents the data, the dot-dashed [green online] line denotes the background, the dotted lines denote the different contributions of each proton feeding, and the dashed [red online] line denotes the total fit. The fit has a $\chi^2_\nu$=1.07. Lower panel: The Residual plot shows the data subtracted from the fit function.}
\label{pie1232-3000}
\end{figure}

 It is important to note that there is very little broadening in the 1232 keV peak due to a long lifetime and therefore any $^{20}$Na states assumed to feed this $^{19}$Ne level yield almost exactly the same peak shape. Therefore, even though the feedings in Lund and Piechaczek differ substantially, they will both fit the data equally well. The lack of sensitivity to the proton branches adopted makes it relatively simple to measure the lifetime of the state. The $\chi^2$ is minimized by taking a value of the lifetime long enough that nearly all the recoiling $^{19}$Ne ions in this state are  stopped before emitting a gamma ray (Fig. \ref{pie1232-3000}). By minimizing the $\chi^2$ as a function of the lifetime, a value of 4.3${^{+1.3}_{-1.1}}$ ps is measured for the lifetime of the 1507 keV state  (Fig. \ref{chisq1232}). The uncertainty is determined from the $\chi^2$ minimization as well as a systematic uncertainty associated with the $\sigma$ parameter and stopping power. This measurement is in agreement with \cite{Gill1970} and more precise, but does not agree within 1 standard deviation with the measurements in \cite{Itahashi1971} or \cite{Tan2005}.
 
The peak at 1269.3 keV is fit using the lifetime of 4.3 ps, determined by the 1232.5 keV peak, since the former peak had much higher statistics. This peak sits next to a Doppler broadened peak from the 1536 keV state $^{19}$Ne that will be addressed in the next section. 

\begin{figure}
\includegraphics[width=0.5\textwidth]{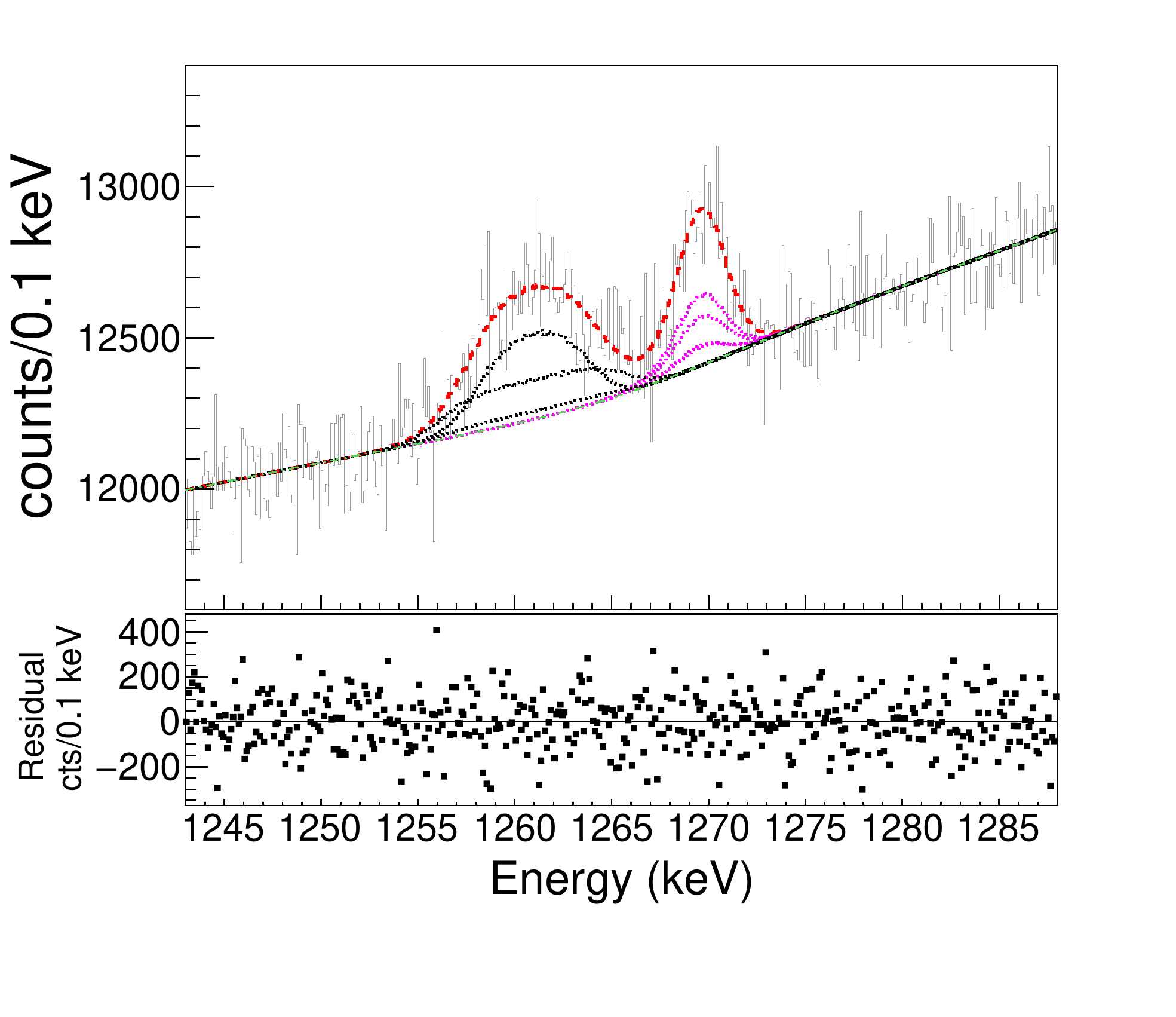}
\caption{(color online) Upper panel: The $\gamma$ ray spectrum above contains two $^{20}$Mg($\beta$p$\gamma$) peaks from different excited states in $^{19}$Ne. The data are represented by the solid [black online] line and, the dot-dashed [green online] line denotes the background, the dotted [pink online] lines denote the different contributions of proton feedings to the 1507 keV state, the dotted [black online] lines denote the different contributions of proton feedings to the 1536 keV state, and the dashed [red online] line denotes the total fit which has a $\chi^2_\nu=1.11$. Lower panel: The Residual plot shows the data subtracted from the fit function.}
\label{pie1269}
\end{figure}

The $\gamma$ ray intensities per $^{20}$Mg $\beta$ decay of the 1232 keV and 1269 keV $\gamma$ rays are shown in Table \ref{tab:intensities}.  We can use these intensities to determine a $\gamma$-decay branching ratio from the 1507 keV state. The uncertainties in efficiency cancel out and we are only concerned with the statistical uncertainty for calculating the branching ratio,  which is measured to be 84.9(4)\% decay to the 275 keV state and 15.1(4)\% decay to the 238 keV state, in agreement with previous measurement \cite{Gill1970}.

The total $\beta$-delayed proton feeding of the 1507 keV state $I_{\beta p-1507}$ = 2.78(7) $\times 10^{-3}$ is consistent with the value from Piechaczek \textit{et al.} of $I_{\beta p-1507}$ = 2.5(3) $\times 10^{-3}$  and more precise but is a factor of 2.7 lower than the value measured by Lund \textit{et al.} $I_{\beta p-1507}$ = 7.4(21) $\times 10^{-3}$, which has a large uncertainty.

\subsection{$^{19}$Ne 1536 keV state}

\begin{table}
\caption{\label{tab:1536}Piechaczek \cite{Piechaczek1995} and Lund \cite{Lund2016} absolute \% proton feeding intensities to 1536 keV state per $^{20}$Mg $\beta$-decay. The quoted uncertainty for all intensities measured by Piechaczek is 12\%.}
\begin{ruledtabular}
\begin{tabular}{@{} cccc @{}}
\multicolumn{2}{c}{\underline{\hspace{11mm}Piechaczek\hspace{9mm}}} & \multicolumn{2}{c}{\underline{\hspace{11mm}Lund\hspace{12mm}}} \\
$E_x(^{20}\text{Na})$ MeV & $I_{\beta p}$ & $E_x(^{20}\text{Na})$ MeV & $I_{\beta p}$\\
\hline
 4.7-5.2 & 0.7  &  & \\
 & & 5.604(5)& 0.03(4)  \\
  6.266(30)&  0.1  &    6.273(7) & 0.33(9)   \\
 6.521(30) & 0.51 &      6.496(3) &  0.47(7)\\
$\approx$6.92&     0.02&   & \\
 $\approx$7.44 &   0.01&   &\\
\end{tabular}
\end{ruledtabular}
\end{table}
There are two $\gamma$ rays which have been measured from this state at 1261 keV and 1298 keV and are expected to have branching ratios of 5(3)\% and 95(3)\% respectively  \cite{Gill1970}. In this work we measure an additional branch decaying to the ground state at 1536 keV for the first time. The lifetime of the state has a recently measured value of 16(4) fs \cite{Tan2005} and is in agreement with the previous evaluation of 28(11) fs \cite{Tilley1995} so a value of 16 fs is adopted for the simulation. Clear broadening is apparent for all three of the $\gamma$ rays emitted and the different proton energies and intensities that feed the 1536 keV state become much more important. For the simulation of each recoil energy, the relatively precise values of $E_x(^{20}\text{Na})$ from Lund \textit{et al.} were adopted. The relative branches from both Piechaczek and Lund were used to separately fit the data and the total number of counts in the peak was left as a free parameter.

\begin{figure}
\includegraphics[width=0.5\textwidth]{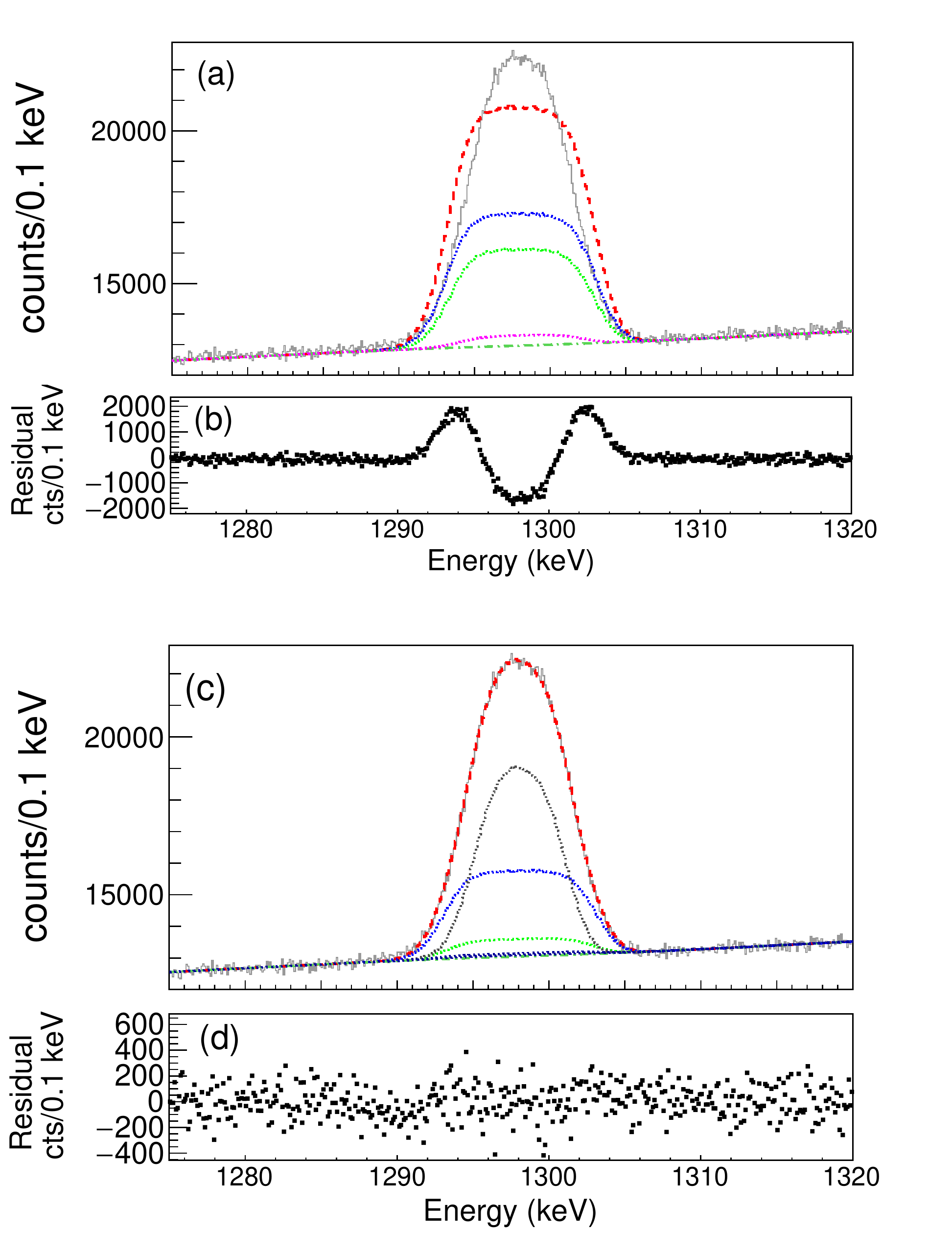}
\caption{(color online) Fits of the 1298 keV $\gamma$ ray peak above are produced using a 16 fs lifetime. (a) The fit is produced using the relative proton feeding intensities, measured by Lund \cite{Lund2016}, from Table \ref{tab:1536}. The data are represented by the solid [gray online] line, the dot-dashed [green online] line denotes the background, the dotted lines denote the different contributions of each proton feeding, and the dashed [red online] line denotes the best total fit. (b) The Residual plot shows the data subtracted from the fit function in (a). (c)  The fit is produced using the relative proton feeding intensities, measured by Piechaczek \cite{Piechaczek1995}, from Table \ref{tab:1536}. Similarly to panel (a) the data are represented by the solid line, the dot-dashed line denotes the background, the dotted lines denote the different contributions of each proton feeding, and the dashed line denotes the best total fit.  (d) The Residual plot shows the data subtracted from the fit function in (c).}
\label{full1298}
\end{figure}

It is easy to see that the relative branches from Lund do not fit the 1298 keV peak accurately with a $\chi^2_\nu$ = 30.8  (Fig. \ref{full1298}). An an additional lower-energy proton feeding is required to fit the data. The relative branches from Piechaczek fit the data much better and return a $\chi^2_\nu$ = 1.14.
From the fit of the 1298 keV peak a value of $I_{\beta p\gamma-1298}$ = ($1.54 \pm 0.03_{stat} \pm 0.03_{sys}$)$\times10^{-2}$ is measured.

A fit of the 1261 keV peak is shown in Fig. \ref{pie1269}. The simulation for this peak used the relative proton feedings from Piechaczek as well as the 16 fs lifetime of the state, which fit the 1298 keV peak well. The feeding of the 1261 keV peak is measured to be  $I_{\beta p\gamma-1261}$ = ($6.75 \pm 0.15_{stat} \pm 0.15_{sys}$)$\times10^{-4}$.

The 1536 keV state has three $\gamma$ decay paths to the ground state of $^{19}$Ne. The two cascades that do not directly decay to the ground state will yield a small portion of counts in the 1536 keV peak due to summing in a single $\gamma$ ray detector. The number of counts in the 1536 keV peak due to the summing effect is calculated from the number of counts in the 1298 keV peak and SeGA efficiency for a 238 keV $\gamma$ ray as well as the number of counts in the 1261 keV peak and SeGA efficiency for a 275 keV $\gamma$ ray. After subtracting the summing counts from the 1536 keV peak integral we measure an intensity of  $I_{\beta p\gamma-1536}$ = ($5.68 \pm 0.44_{stat} \pm 0.17_{sys}$)$\times10^{-4}$.

From $I_{\beta p\gamma-1261}$, $I_{\beta p\gamma-1298}$, and the newly measured $I_{\beta p\gamma-1536}$  we measure the $\gamma$ ray branching ratio from the $^{19}$Ne 1536 keV state to be a 4.05(16)\% branch to the 275 keV state, a 92.53(35)\% branch to the 238 keV state, and a 3.42(29)\% branch to the ground state.

\subsection{$^{19}$Ne 1615 keV state}
There are three $\gamma$ rays which are emitted from this state with energies of 1340, 1377, and 1615 keV and they are expected to have branching ratios of 70(4)\%, 10(3)\%, and 20(3)\% respectively \cite{Gill1970}. This state has never been observed in $^{20}$Mg $\beta$-decay before the present work, so there is no available proton feeding data. It is possible that multiple $^{20}$Na states contribute to the feeding, however, the simplest procedure is to begin by assuming one proton energy to fit the peak and this CoM energy will be considered a free parameter. A lifetime of 143(31) fs was determined in a data evaluation \cite{Tilley1995} by combining measurements from \cite{Gill1970,Lebrun1977}, however a more recent value of 80(15) fs was reported \cite{Tan2005} so we have re-evaluated the lifetime to be 93(20) fs by taking a weighted average with inflated uncertainty.

\begin{figure}
\includegraphics[width=0.5\textwidth]{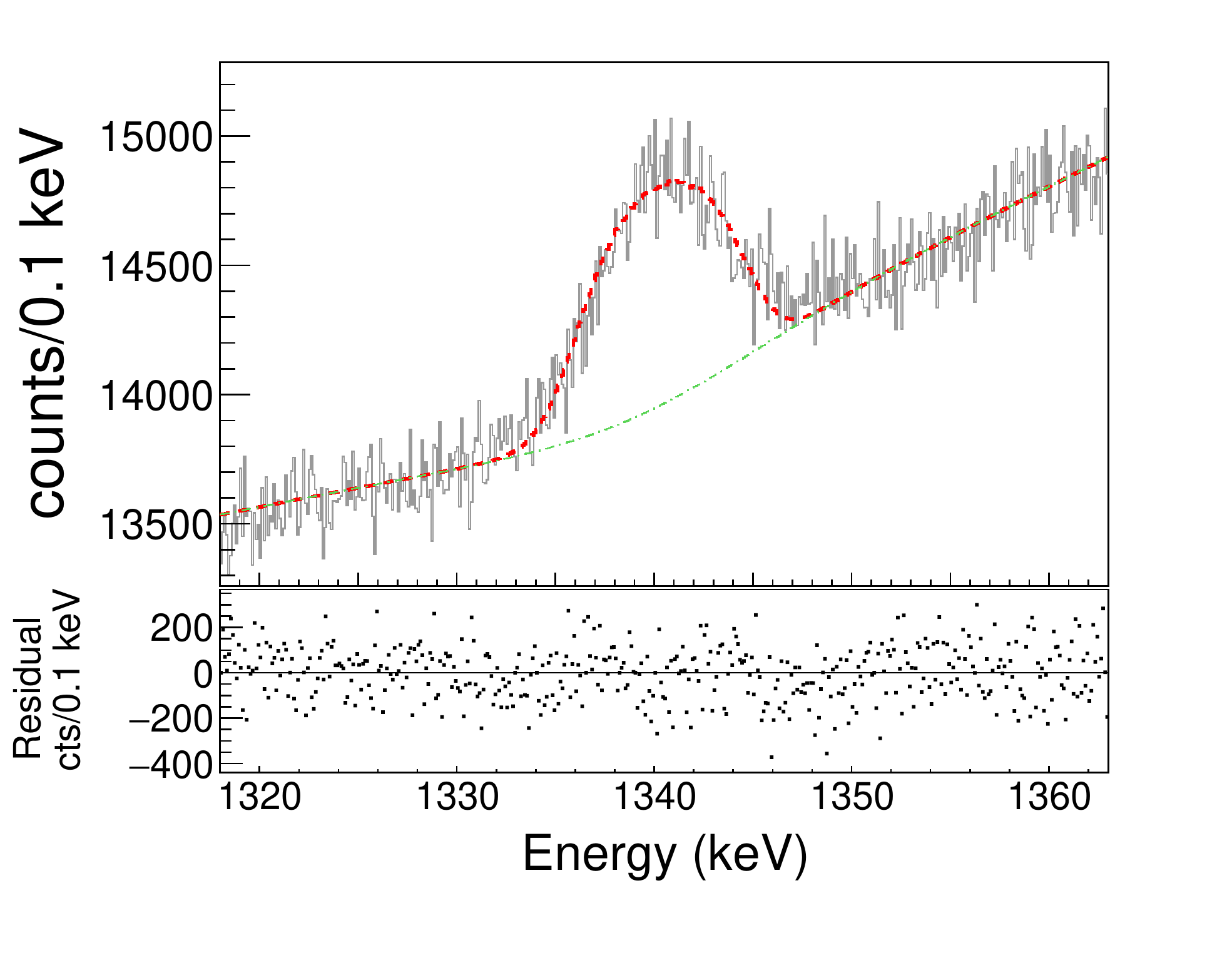}
\caption{(color online) Upper panel: The fit of the 1340 keV $\gamma$ ray peak is produced by using a 93 fs lifetime and a CoM energy of 2.7 MeV between the proton and recoiling $^{19}$Ne. The solid [gray online] line represents the data, the dot-dashed [green online] line denotes the background and the dashed [red online] line denotes the background+simulated peak. The fit has a $\chi^2_\nu$=1.00. Lower panel: The Residual plot shows the data subtracted from the fit function.}
\label{fit1340}
\end{figure}

 Using the adopted lifetime of 93(20) fs and interpolated $\sigma$ parameter to simulate the broadening of the 1340 keV peak, a CoM energy of 2.7 MeV minimizes the $\chi^2$ (Fig. \ref{fit1340}). From the $\chi^2$ distribution we get an uncertainty in the CoM energy of 100 keV. An additional systematic uncertainty in the CoM energy of 200 keV from the uncertainty in the lifetime as well as an uncertainty of 50 keV for the uncertainty in the $\sigma$ parameter yields a value of 2.70(23) MeV for the CoM energy. From this we determine an excitation energy $E_x(^{20}\text{Na})$= 6.51(23) MeV for the proton-emitting state. This is consistent with proton emission from the $^{20}$Na isobaric analog state at 6498.4(5) keV \cite{Glassman2015}.
 
 The 1377 and 1615 keV lines both have low statistics and do not provide significant information about the energies of protons feeding the state. We apply the peak shape corresponding to the proton energies that best fit the higher statistics 1340 keV peak to these two peaks to determine the total intensity of protons feeding the 1615 keV state. 

For a fit of the 1377 keV peak a simple linear background was used for this relatively low statistics case and a broad peak was fit on top of it. In the case of the 1615 keV peak a linear plus exponential function was used to model the background since the peak sits on the tail of a very high statistics 1634 keV peak from $^{20}$Na($\beta\gamma$) decay. 

Since the 1615 keV state also has two cascades that do not directly decay to the ground state, a small portion of counts in the 1615 keV peak are due to summing in a single $\gamma$ ray detector and must be subtracted. The number of counts in the 1615 keV peak due to this effect is calculated from the number of counts in the 1340 keV peak and SeGA efficiency for a 275 keV $\gamma$ ray as well as the number of counts in the 1377 keV peak and SeGA efficiency for a 238 keV $\gamma$ ray.

The $\gamma$ ray intensities per $^{20}$Mg $\beta$ decay of the 1340, 1377, and 1615 keV $\gamma$ rays are shown in Table \ref{tab:intensities}.
A measurement of the branching ratios from the 1615 keV state using the intensities yields a 74.0(17)\% branch to the 275 keV state, a 8.6(18)\% branch to the 238 keV state, and a 17.4(9)\% branch to the ground state of $^{19}$Ne, in agreement with and more precise than previous measurement \cite{Gill1970}.

%

\subsection{$^{19}$Ne 4.03 MeV state}
\begin{figure}
\includegraphics[width=0.5\textwidth]{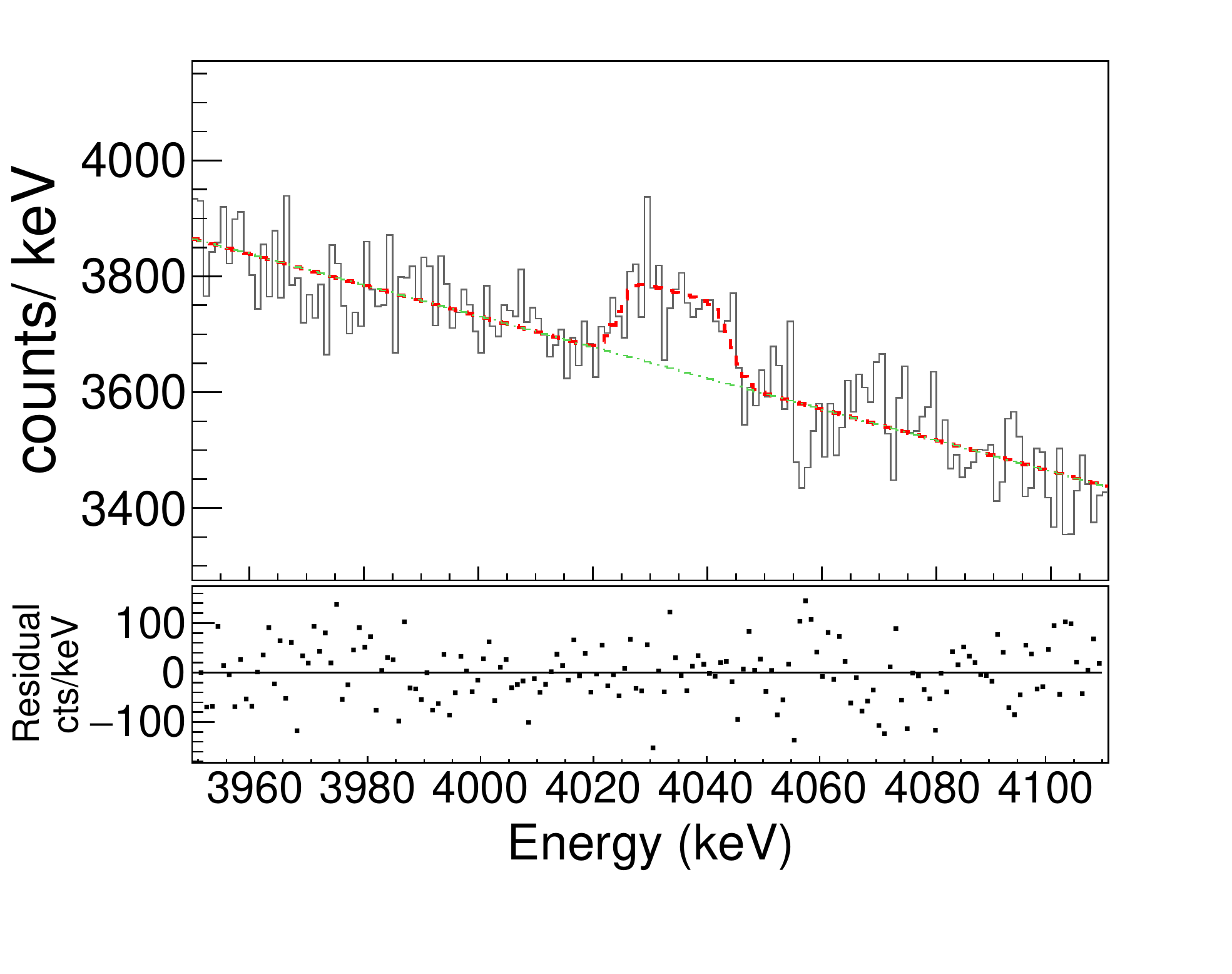}
\caption{(color online) Upper panel:  The fit of the 4.03 MeV peak is produced by simulating the broadened peak with an 7 fs lifetime and CoM energy of 1.21 MeV and has a $\chi^2_\nu=0.94$. All 16 SeGA detectors are used to produce this spectrum. The solid [gray online] line represents the data, the dot-dashed [green online] line denotes a fit of the background and the dashed [red online] line denotes the total fit using the optimal 1.21 MeV CoM energy. A simplified linear background model was applied for this relatively low statistics case. 
Lower panel: The Residual plot shows the data subtracted from the fit function.}
\label{fit4033}
\end{figure}

There are three $\gamma$ rays which are emitted from this state at 2497, 3758, and 4034 keV and they are expected to have branching ratios of 15(5)\%, 5(5)\%, and 80(15)\% respectively \cite{Tilley1995}. In the present experiment, only the 4.03 MeV $\gamma$ ray is detected above background. For this case all 16 detectors are used to determine the feeding of the 4.03 MeV state and the shape of the Doppler broadened feature in order to reduce the statistical uncertainty.

The lifetime of the 4.03 MeV state has been measured to be 13${^{+16}_{-9}}$ fs \cite{Tan2005}, 11${^{+4}_{-3}}$ fs \cite{Kanungo2006}, and 6.9$\pm1.7$ fs \cite{Mythili2008}. The more precise lifetime of 6.9 fs was adopted and the uncertainty is used to determine a systematic uncertainty in the CoM energy which was left as a free parameter. In this case, where the statistics are relatively low, a simple linear model was used for the background. Additionally, an assumption is made that only one $^{20}$Na excited state feeds the 4.03 MeV level (Fig. \ref{fit4033}). 

\begin{figure}
\includegraphics[width=0.5\textwidth]{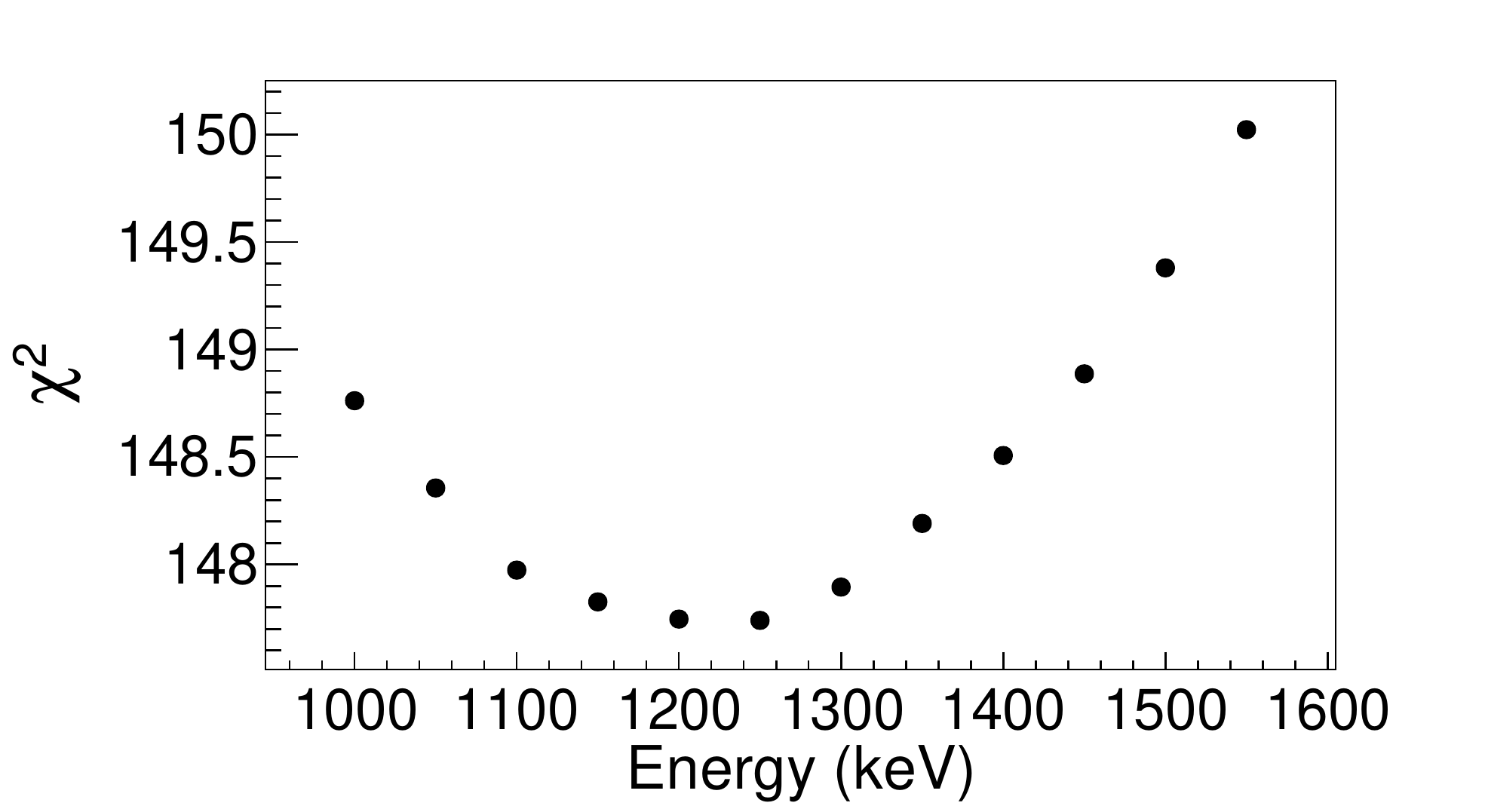}
\caption{Each $\chi^2$ value is determined by simulating a different CoM proton energy feeding the 4.03 MeV excited state in $^{19}$Ne and comparing each simulation to the peak at 4034 keV. $\chi^2$ values are determined from fits with 157 degrees of freedom. The minimum determines the most likely CoM energy.}
\label{chisq4033}
\end{figure}

Minimizing the $\chi^2$ as a function of CoM energy (Fig. \ref{chisq4033}) yields a CoM energy of 1.21${^{+0.25}_{-0.22}}$ MeV. An additional 0.025 MeV is incorporated into this uncertainty from the shift in minimum $\chi^2$ introduced by moving the lifetime to the limits of uncertainty. This corresponds to a feeding from an excited state in $^{20}$Na at 7.44 ${^{+0.25}_{-0.22}}$ MeV, consistent with the 7.44(10) MeV state observed to be populated in $^{20}$Mg $\beta$-decay by its proton emission to lower lying $^{19}$Ne states \cite{Piechaczek1995}.

From this fit the intensity is measured to be $I_{\beta p\gamma-4034}$ =  ($1.19 \pm 0.12_{stat} \pm 0.12_{sys}$)$\times10^{-4}$. The $\gamma$ branch from the 4.03 MeV state is expected to be 80(15)\% \cite{Tilley1995}. Therefore, $I_{\beta p-4034}$ =  ($1.49 \pm 0.15_{stat} \pm 0.32_{sys}$)$\times10^{-4}$. This value is consistent with the one reported in \cite{Wrede2017} but slightly different because the fitting procedure is different and a different literature intensity was adopted for the 984-keV $^{20}$Na line for normalization.

\subsection{$^{19}$Ne 238 and 275 keV states}
Both of these lower lying $^{19}$Ne states have long lifetimes, and the corresponding $^{19}$Ne atoms are completely stopped in the scintillator before emitting $\gamma$ rays. Therefore, we do not gain any information from Doppler broadening analysis. However, the direct feeding of the 238 and 275 keV states from $^{20}$Mg($\beta p$) decay can be determined by measuring the intensity of the $\gamma$ decays and subtracting the feeding contribution to each of these states from $\gamma$ decays of higher lying states in $^{19}$Ne. Both of these states are fed by the 1507, 1536, 1615 and 4034 keV states and these contributions are subtracted to obtain the intensities reported in Table \ref{tab:intensities}. These values are consistent with the previously measured values of $I_{\beta p-238}$ = ($2.29 \pm 0.27$)$\times10^{-2}$ and $I_{\beta p-275}$ = ($3.12 \pm 0.37$)$\times10^{-2}$ \cite{Piechaczek1995} and $I_{\beta p-238}$ = ($2.23 \pm 0.34$)$\times10^{-2}$ and $I_{\beta p-275}$ = ($3.69 \pm 0.52$)$\times10^{-2}$ \cite{Lund2016} and more precise.
\\

\section{Conclusion}
We have measured the $^{20}$Mg($\beta p$)$^{19}$Ne feedings and $\gamma$ ray branches of 6 excited states in $^{19}$Ne. We have developed a Monte Carlo simulation to analyze 9 Doppler broadened $^{19}$Ne peaks. We have measured the energy of the proton transition which feeds the astrophysically important 4.03 MeV state, facilitating future measurements of the $\alpha$-branch from this state. Additionally we have measured the energy of the proton transition which feeds the 1615 keV state as well as the lifetime of the 1507 keV state and found a new $\gamma$ decay branch from the 1536 keV state.

This is the first time Doppler broadening analysis has been applied to such high statistics $\beta$-delayed proton-$\gamma$ peaks, enabling a substantial improvement in sensitivity over \cite{Schwartz2015}. We have shown this method can be a useful tool to measure excited state lifetimes, proton branches, and proton energies and can distinguish between conflicting decay schemes. The method is therefore complementary to direct measurements of $\beta$-delayed protons and should prove to be even more useful when applied to $\beta$-delayed neutron emission.

\section{Acknowledgements}
We gratefully acknowledge the NSCL staff for technical assistance and for providing the $^{20}$Mg beam. This work was supported by the National Science Foundation (USA) under Grants No. PHY-1102511, No. PHY-1419765, No. PHY-1404442, and No. PHY-1430152, the U.S Department of Energy, Office of Science, under Award No. DE-SC0016052, Contract No. DE-AC05-00OR22725, and the U.S. Department of Energy National Nuclear Security Administration under Awards No. DE-NA0003221 and No. DE-NA0000979.


\bibliographystyle{apsrev4-1}
\bibliography{DopplerBbibl}
\end{document}